\documentclass[twocolumn,amsmath,nofootinbib,amssymb,showpacs]{revtex4}

\usepackage{multirow}
\usepackage[english]{babel}
\usepackage{amsmath}
\usepackage{amssymb}
\usepackage{epsfig}
\usepackage{graphics,psfrag,rotating}
\usepackage{graphicx}
\usepackage{dcolumn}
\usepackage{bm}
\usepackage{slashbox}
\usepackage{array,multirow}
\usepackage{epsfig}
\usepackage{graphicx}
\usepackage{dcolumn}
\usepackage{bm}
\usepackage{color}

\newcommand{\lsim}{\mbox{\raisebox{-.9ex}{~$\stackrel{\mbox{$<$}}{\sim}$~}}}

\def\thebiblio#1{
\begin{center}\bf \large References
\end{center}
\list
{[\arabic{enumi}]}{\settowidth\labelwidth{#1.}\leftmargin\labelwidth
 \advance\leftmargin\labelsep
 \usecounter{enumi}}
 \def\newblock{\hskip .11em plus .33em minus -.07em}
 \sloppy
 \sfcode`\.=1000\relax}

\begin{document}
\preprint{}
\title{%
Neutrino fluxes from Dark Matter in the HESS J1745-290 source\\
at the Galactic Center}

\author{J. A. R. Cembranos \footnote{E-mail:cembra@fis.ucm.es},
V. Gammaldi \footnote{E-mail:vivigamm@pas.ucm.es},
and A.\,L.\,Maroto \footnote{E-mail:maroto@fis.ucm.es}
}

\affiliation{Departamento de  F\'{\i}sica Te\'orica I, Universidad Complutense de Madrid, E-28040 Madrid, Spain}%
\date{\today}

\begin{abstract}
The spectral study of the HESS J1745-290 high energy gamma-ray cut-off from the galactic center is compatible with a signal of
Dark Matter (DM) annihilation or decay. If this is the case, a neutrino flux from that source is also expected. We analyze the neutrino
flux predicted by DM particles able to originate the HESS J1745-290 gamma-rays observations. We focus on the electroweak and hadronic
channels, which are favoured by present measurements. In particular, we study DM annihilating into $W^+W^-$ and $u\bar u$
with DM masses of $48.8$ and $27.9$ TeV respectively. We estimate the resolution angle and exposition time necessary to
test the DM hypothesis as the origin of the commented gamma signal.
\end{abstract}

\maketitle

\section{Introduction}

Different telescopes have observed Very High Energy (VHE) gamma-rays coming from the Galactic Center (GC), such as
CANGAROO \cite{CANG}, VERITAS \cite{VER}, MAGIC \cite{MAG} or Fermi-LAT  \cite{Vitale, ferm}.
In this work, we will pay attention to the data collected by the HESS collaboration from the J1745-290 source
during the years 2004, 2005, and 2006 \cite{Aha, HESS}. The variability of the IR and X-ray observations \cite{X}
indicates a different emission mechanism for this part of the spectrum. In addition, one of the most
characteristic features of the HESS J1745-290 data consists in a cut-off at several tens of TeVs. These spectral properties
can be explained naturally by the photons produced by the annihilation or decay of Dark Matter (DM) particles.
This interpretation was discussed from the very early days of the publication of the HESS data \cite{Bergstrom1,DMint} but
it was concluded that the DM origin was disfavored \cite{DMint}. However, a recent study has shown that the observed data are
well fitted as DM signal complemented by a diffuse background \cite{HESSfit}. Indeed, this background has a
good motivation since VHE photons are also expected from radiative processes generated by particle acceleration in the
neighborhood of the supermassive black hole Sgr A and the Sgr A East supernova. The analysis shows good agreement with DM annihilation
or decay into $u\bar u$, $d\bar d$, $s\bar s$ and $t\bar t$ quark-antiquark channels and $W^+W^-$ and $ZZ$ boson channels.
Leptonic and other quark-antiquark channels were excluded with $95.4\%$ confidence level. The background provided by the analysis
is also compatible with the Fermi-LAT data from the IFGL J1745.6-2900 source observed during 25 months \cite{ferm}, which is spatially
consistent with the HESS J1745-290 source \cite{Cohen}.

In any case, the fundamental nature of this gamma-ray flux is still unclear. The entire VHE spectrum may be produced by particle propagation
\cite{ferm,SgrA} in the vicinity of the commented supernova remnant and black hole, both located at the central region of our
galaxy \cite{Atoyan,AN}. In addition, the emission region is quite compact since the signal is limited to a region of few tenths
of degree \cite{HESS}. This feature is not consistent with dark halos simulated with non-baryonic cold DM, such as the standard
NFW profile \cite{Navarro:1996gj}. It needs to be more compact as the ones produced when baryonic effects are taken
into account. It has been argued that the baryonic gas falls to the inner part of the halo, modifying the gravitational potential
and increasing the DM density in the center \cite{Blumenthal,Prada:2004pi}. This scenario is not completely accepted (read \cite{Romano}
for example), but if it is correct, it has two important consequences. First, the sensitivity of indirect DM searches is reduced to a
more compressed region; and second, the DM annihilating fluxes are enhanced by up to three orders of magnitude with respect to the standard
NFW profile \cite{Prada:2004pi}. The HESS observations are in good agreement with these types of compressed dark halos.

The DM particle that originate this spectrum needs to have a mass between $15 \; \text{TeV} \lesssim M \lesssim 110 \; \text{TeV}$ \cite{HESSfit}.
This makes highly challenging to observe these particles in direct detection experiments or particle accelerators \cite{lab}. On the contrary,
complementary cosmic rays analysis \cite{cosmics} from the GC and from other astrophysical objects are the most promising way to cross check the
commented DM hypotheses. In particular, the analysis of neutrino fluxes from the same region can be determining.
If DM annihilates or decays into Standard Model (SM) particles producing VHE gamma-rays photons, it has to produce also VHE neutrinos. Indeed, if the
dark halo properties are adjusted to explain the HESS J1745-290 data, the neutrino flux is completely determined if one concrete annihilation
or decay channel is assumed.
This work is organised as follows: In Section II, we study the expected neutrino fluxes as indirect products of annihilating DM in the direction of the
GC. Section III is devoted to discuss the flavor oscillation effects in this signal. In Section IV, we model the background of our analysis
by taking into account the atmospheric neutrino flux observed by the IceCube experiment and we study the best configuration that may allow
the detection of the corresponding neutrino signal associated with the HESS J$1745$-$290$ GC gamma-rays source. Finally, we summarize our main
conclusions in Section V.

\begin{figure}[bt]
\begin{center}
\epsfxsize=13cm
\resizebox{8.8cm}{6.6cm}
{\includegraphics{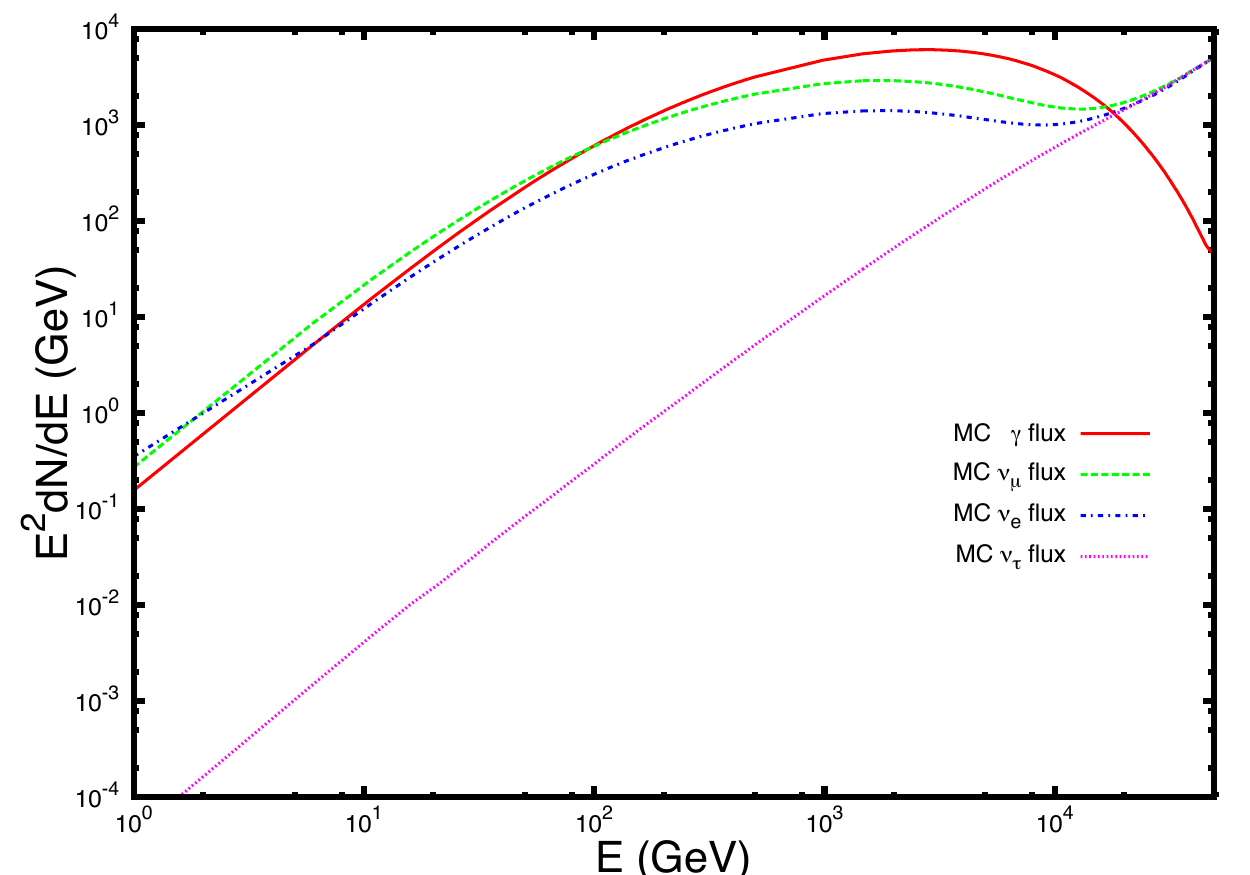}}
\caption {\footnotesize{ The gramma-ray ($\gamma$) and neutrino ($\nu_p$) fluxes from DM annihilating into $W^+W^-$ bosons,
as generated by PYTHIA 8.135 and reported by \cite{Cirelli}.} }
\label{MCfluxes}
\end{center}
\end{figure}

\section{Astrophysical neutrino flux}

The differential flux of neutrinos of a given flavor $\nu_f$ observed on the Earth in a particular direction can be computed as
\begin{eqnarray}
\frac{d \Phi_{\nu_f}}{dE} &=&\sum^3_{p=1}\sum^2_{a=1} \sum^{\text{channels}}_i P_{fp} \cdot \frac{\zeta^{(a,\,\nu_p)}_{i}}{a}
\nonumber\\
&& \frac{dN^{(a,\,\nu_p)}_{i}}{dE}
 \cdot \frac{{\Delta\Omega\,\langle J_{(a)} \rangle}_{\Delta\Omega}}{4 \pi M^a}\,,
\label{nuflux}
\end{eqnarray}
where $P_{fp}$ are the elements of the symmetric $3\times 3$ matrix which takes into account the neutrino oscillation effects
from the produced neutrino flavor ($\nu_p$) generated by the DM from galactic sources to the observed neutrino flavor ($\nu_f$) on the
Earth. We shall discuss these effects in detail in the next section. $M$ is the mass of the DM particle.
The case $a=2$ accounts for neutrinos coming from DM annihilation with $\zeta^{(2,\,\nu_p)}_i\equiv\langle\sigma^{\nu_p}_i v\rangle$
the thermal averaged annihilation cross-section of two DM particles (assumed to be their own antiparticles) into SM particles
(also labeled by the subindex i). If DM is meta-stable, neutrinos can be produced also by its decay. In such a case the contribution with
$a=1$ is activated with $\zeta^{(1,\,\nu_p)}_i\equiv\Gamma^{\nu_p}_i$ the decay width into SM particles (labeled by the same subindex $i$).

The number of neutrinos of flavor $\nu_p$ produced in each annihilating or decaying channel $dN^{(a,\,\nu_p)}_{i}/dE$,
involves decays and/or hadronization of unstable products such as quarks and leptons. Because of the non-perturbative QCD effects, this
requires Monte Carlo events generators \cite{pythia} or fitting or interpolation functions \cite{Ce10}. In particular, we will use
the results reported in \cite{Cirelli}. They refer to Pythia 8.135 Monte Carlo events generator software \cite{pythia} and reproduce the differential number of neutrinos produced by DM of different masses. In this work, we will focus on neutrino fluxes coming from fragmentation and decays of SM particle-antiparticle pairs produced by DM annihilation. We shall ignore DM decays, the possible production of mono energetic neutrinos, n-body annihilations (with $n>2$), or neutrinos produced from electroweak bremsstrahlung. In particular, we will consider DM annihilation into single channels of SM particle-antiparticle pairs
that are consistent with the origin of the HESS J$1745$-$290$  gamma-ray observations as we have explained.

\begin{figure}[bt]
\begin{center}
\epsfxsize=13cm
\resizebox{8.8cm}{6.6cm}
{\includegraphics{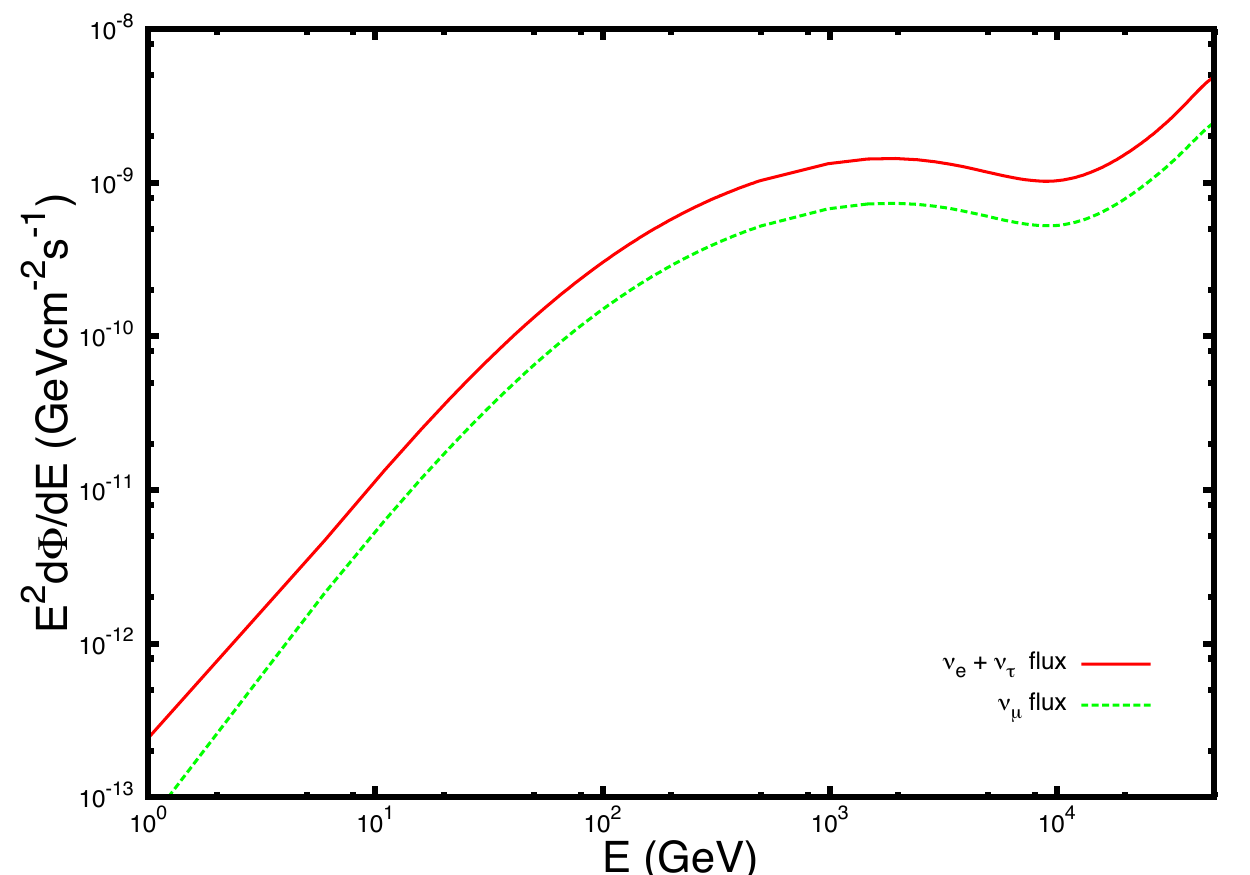}}
\caption {\footnotesize{Neutrino differential fluxes ($\Phi_{\nu_e}+\Phi_{\nu_\tau}$ and $\Phi_{\nu_\mu}$) as expected to be observed on the Earth,
taking into account both neutrinos oscillations and neutrino-antineutrino total flux. We are assuming DM annihilating into the $W^+W^-$
channel. The parameters in Eq.(\ref{nuflux}) are: $\langle\sigma v\rangle=3\times10^{-26}\text{cm}^{3}\text{s}^{-1}$, $M=48.8$ TeV, $<J_{(2)}>_{\Delta\Omega}\simeq 4.95\times10^{28}\text{GeV}^2\text{cm}^{-5}$, and $\Delta\Omega=10^{-5}$.}}
\label{EXPflux}
\end{center}
\end{figure}

The DM spatial distribution is encoded in the astrophysical factors $\langle J_{(a)} \rangle$, that depend on the $\Psi$ angle, determined by
the line of observation with respect to the direction of the GC, and the total angular field of view $\Delta\Omega$:
\begin{eqnarray}
\langle J_{(a)} \rangle= \frac{1}{\Delta\Omega}\int_{\Delta\Omega}\text{d}\Omega\int_0^{l_{max}(\Psi)} \rho^a [r(l)] dl(\Psi)\,,
\label{astrofactor}
\end{eqnarray}
where $l$ is the distance from the Sun to a particular point of the DM halo, that is related to the radial distance $r$, computed respect to the GC,
through the equation: $r^2 = l^2 + D_\odot^2 -2D_\odot l \,\cos \Psi$. The distance between the Sun and the center of the Galaxy is denoted by
$D_\odot \simeq 8.5$ kpc, and the maximum distance between the Sun and the edge of the halo in a given direction $\Psi$ is
$l_{max} = D_\odot \cos \Psi + \sqrt{r_{max}^2-D_\odot^2 \sin \Psi}$. The differential astrophysical factor is proportional to $\rho^2$ when
it accounts for DM annihilation and proportional to $\rho$ when it computes a DM decay.

As we have commented, the neutrino fluxes have to be averaged over the field of view of the detector, that we shall parameterize
with the angle $\theta$: $\Delta \Omega = 2 \pi ( 1 - \cos \theta )$. The HESS Cherenkov telescopes array can be characterized
typically by $\Delta \Omega_{\text{HESS}}\simeq 10^{-5}$ or $\theta_{\text{HESS}} \simeq 0.1^\circ$. This angular resolution angle
is not precise enough to resolve the J$1745$-$290$  gamma-ray morphology, which can be approximated by a point-like source. Therefore,
the integration along the line of sight can be approximated by a constant value for $\theta \gtrsim 0.1^\circ$ and the astrophysical factor given
by Eq. (\ref{astrofactor}) is fixed by fitting the HESS data:

\begin{figure}[bt]
\begin{center}
\epsfxsize=13cm
\resizebox{8.8cm}{6.6cm}
{\includegraphics{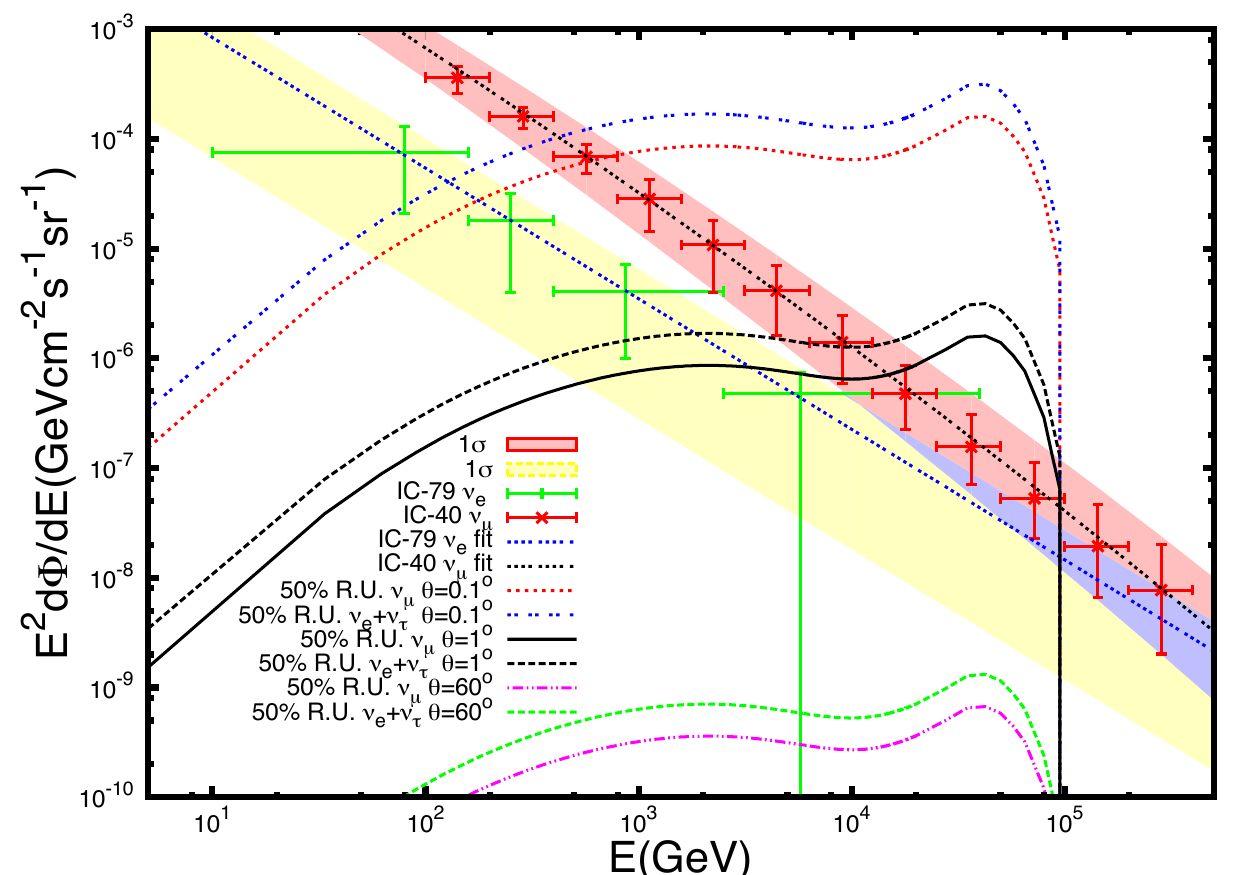}}
\caption {\footnotesize{
Expected neutrino fluxes corresponding to muon neutrinos and electron plus tau neutrinos
from DM annihilating into $W^+W^-$ bosons for an angular field of view of $\theta=60^\circ$, $1^\circ$ and $0.1^\circ$.
The flux accounts for a $50\%$ Resolution Uncertainty (R.U.) associated with a typical high energy neutrino telescope.
The observed atmospheric muon by the IceCube telescope in the 40-string configuration (IC-40) and electron neutrinos by the 79-string configuration (IC-79)
are also shown together with the fitting functions given by Equations (\ref{Bgmu}) and (\ref{Bge})
respectively and the corresponding shared regions at $1\sigma$ confidence level.
}}
\label{WFigRIS}
\end{center}
\end{figure}

\begin{equation}
\langle J_{(a)} \rangle= \langle J_{(a)} \rangle_\text{HESS}\,\frac{\Delta \Omega_{\text{HESS}}}{\Delta \Omega}\,,
\label{astroneutrino}
\end{equation}
where $\langle J_{(a)} \rangle_\text{HESS}$ is the astrophysical factor which reproduces the J$1745$-$290$  gamma-ray flux, and
it depends on the particular annihilating or decaying DM channel \cite{HESSfit}.
Therefore, for a neutrino telescope with $\Delta \Omega \gtrsim 10^{-5}$ the total astrophysical factor ($\langle J_{(a)} \rangle \Delta \Omega$)
is constant, whereas the average ($\langle J_{(a)} \rangle$) decreases with $\Delta \Omega$ inversely. In particular, we will focus on the
$W^+W^-$ and $u\bar u$ annihilation channels with the standard thermal value $\langle\sigma v\rangle=3\times10^{-26}\text{cm}^{3}\text{s}^{-1}$.
By taking into account the results of \cite{HESSfit}:
\begin{equation}
\langle J^{W^+W^-}_{(2)} \rangle= \frac{(7.9\pm 1.9) \times 10^{-22}}{1 - \cos \theta}\; \text{GeV}^2\text{cm}^{-5}\,,
\label{astroneutrinoWW}
\end{equation}
and
\begin{equation}
\langle J^{u\bar u}_{(2)} \rangle= \frac{(4.4\pm 0.8) \times 10^{-22}}{1 - \cos \theta}\; \text{GeV}^2\text{cm}^{-5}\,.
\label{astroneutrinouub}
\end{equation}


\begin{figure}[bt]
\begin{center}
\epsfxsize=13cm
\resizebox{8.8cm}{6.6cm}
{\includegraphics{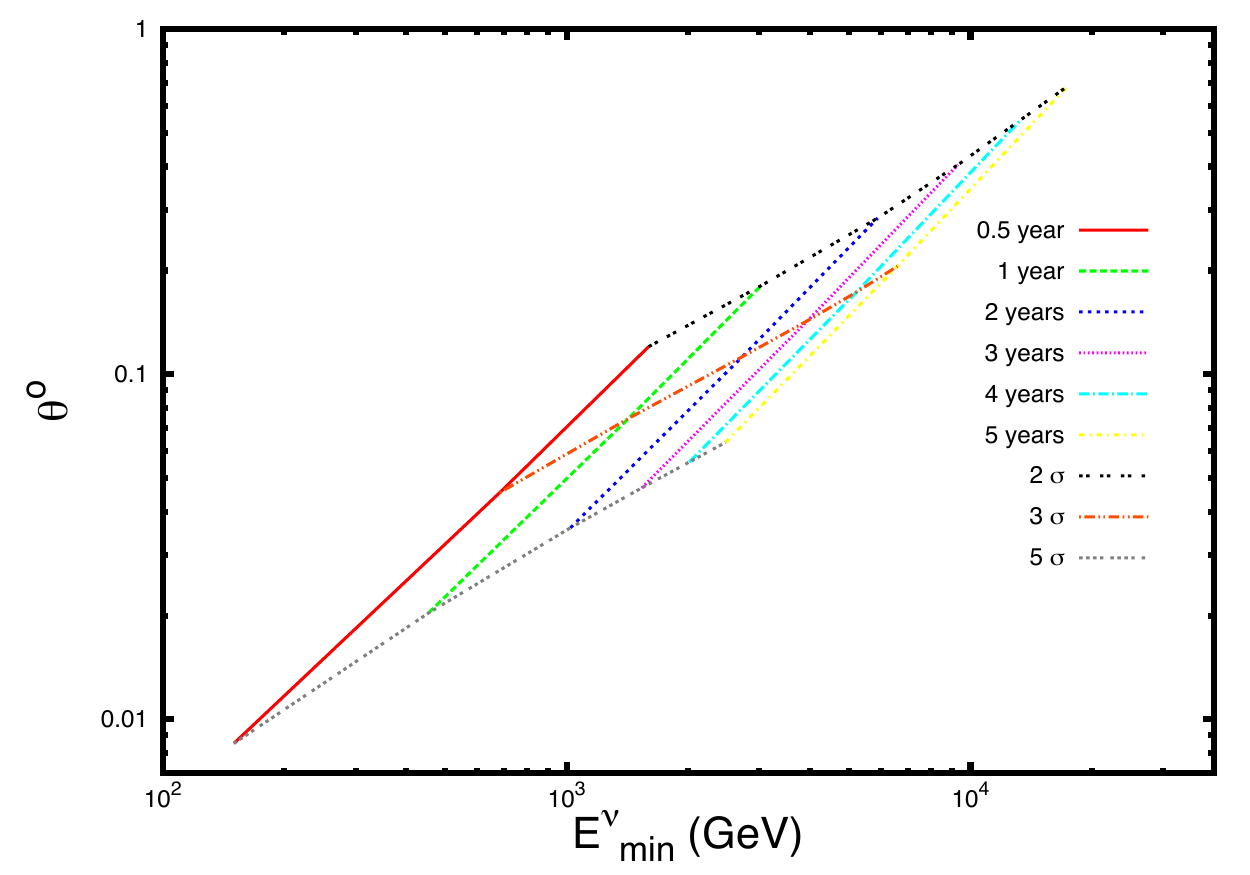}}
\caption {\footnotesize{Combination of the angular field of view $\theta$, minimum energy threshold and exposition time that allow to detect a muon neutrino
flux signal coming from DM annihilating in the GC at $2\sigma$, $3\sigma$ or $5\sigma$ confidence level, with a detector with $50\;\text{m}^2$ effective area.  The annihilating mode is the $W^+W^-$ channel,
the mass of the DM particles is $48.8$ TeV, the annihilation cross section is $\langle\sigma v\rangle=3\times10^{-26}\text{cm}^{3}\text{s}^{-1}$ and the
astrophysical factor is given by Eq. (\ref{astroneutrinoWW}). The lowest value of $\theta\simeq 0.01^\circ$ corresponds to a $5\sigma$ confidence level
with energy threshold of $E^\nu_{\text{min}}\simeq 150$ GeV and six months of exposition time. The higher the exposition time, the higher the angular
resolution of the analysis needed to reduce the atmospheric background. The largest value of $\theta\simeq 0.68^\circ$ is associated with $5$ years of
exposition time, a statistical significance of $2\sigma$, and an energy threshold of $E^\nu_{\text{min}}\simeq 17.42$ TeV.}}
\label{tfixRIS}
\end{center}
\end{figure}

\section{Neutrino flavors and mixing}
After simulating the neutrino fluxes produced at the source, one has to take into account different aspects in order to estimate the expected flux as observed on the Earth, such as neutrino oscillations and detector sensitivity to neutrino flavors. On the other hand, we shall assume that our detector is not able to discriminate between neutrinos and antineutrinos. Due to neutrino oscillations, the ratio of neutrino flavor changes during the way from the source to the observer \cite{NeutrinoOsci}. By considering the standard three-flavor neutrino oscillation, the probability matrix $P$ for astrophysical neutrinos traversing a vast distance is given by:
\begin{equation}
P(i\rightarrow j)=\sum_{a=1}^3 |U_{ia}|^2|U_{ja}|^2,
\end{equation}
where $U_{ia}$ are the elements of the neutrino mixing matrix \cite{Lai}. For example, for the simplified case of the oscillation between only two flavors at distance $x$ by the source, the probability can be written as:
\begin{equation}
P(i\rightarrow j)=\text{sin}^2(2\alpha_{ij})\times\text{sin}^2\left(\pi\frac{x}{L}\right).
\end{equation}

\begin{figure}[bt]
\begin{center}
\epsfxsize=13cm
\resizebox{8cm}{8cm}
{\includegraphics{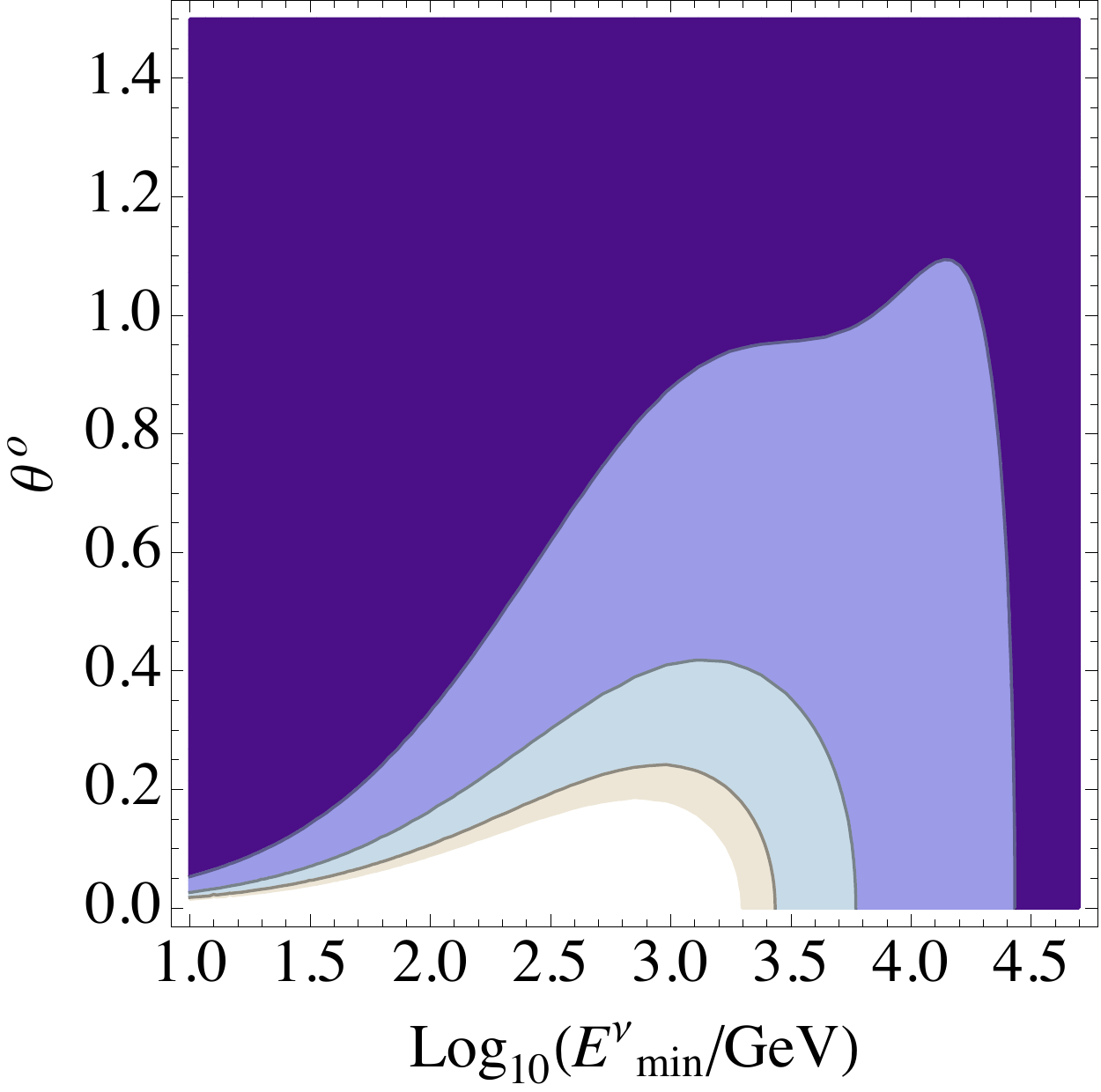}}
\end{center}
\begin{center}
\epsfxsize=13cm
\resizebox{8cm}{8cm}
{\includegraphics{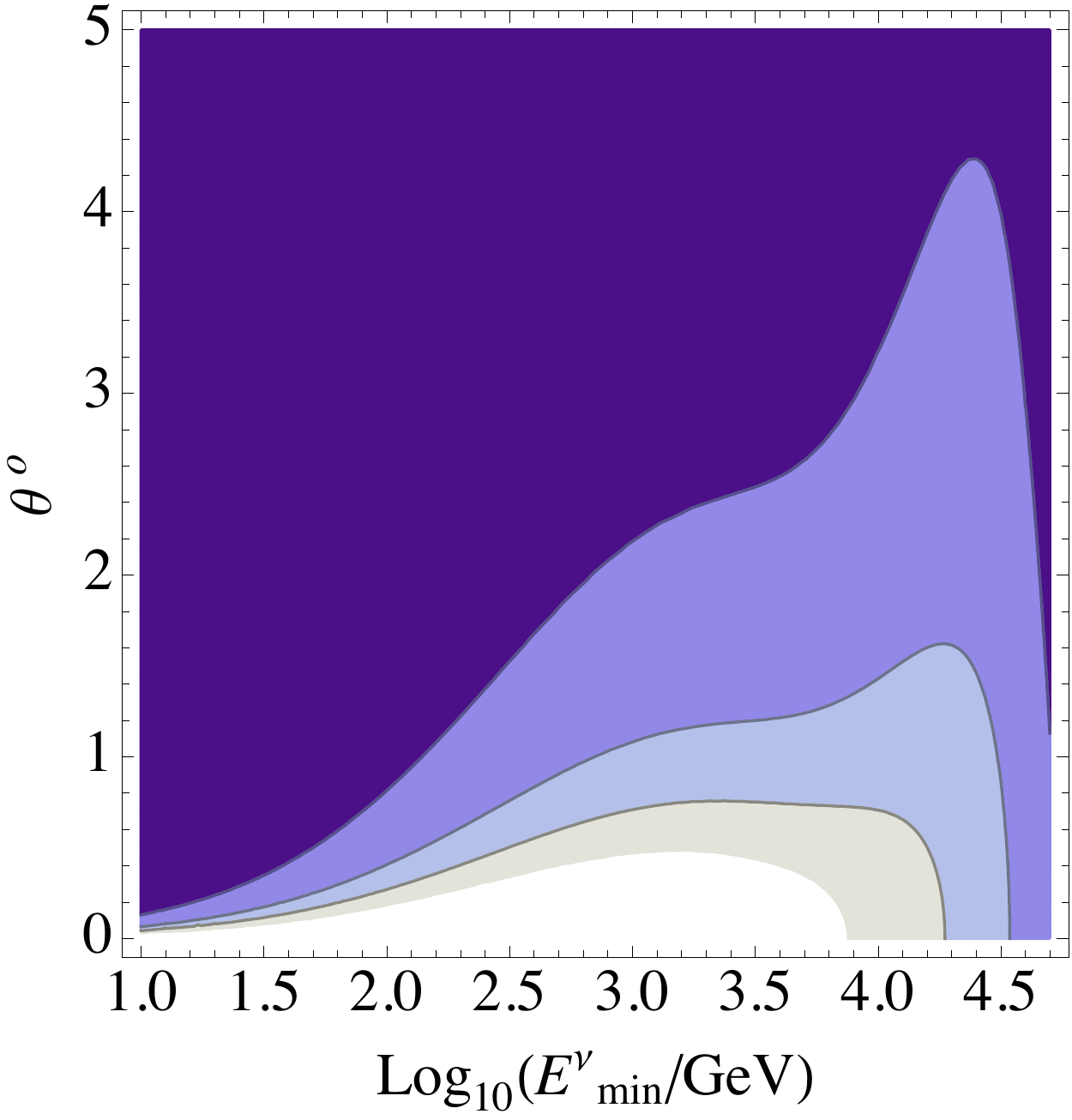}}
\caption {\footnotesize{The Figure shows the $1\sigma$ (dark), $2\sigma$, $3\sigma$, $5\sigma$ (white) confidence levels contours
in the case of DM annihilating into the $W^+W^-$ channel. The factor $Af=A_{\text{eff}}\times t_{\text{exp}}$ is fixed in both analyses:
$Af=100\, \text{m}^2\,\text{yr}$ (top pannel) and $Af=600\, \text{m}^2\,\text{yr}$ (bottom pannel). The possibility to detect the neutrino flux signal
above the atmospheric background depends on the energy cut $E_{\text{min}}^\nu$ and the resolution angle. }}
\label{thetaW}
\end{center}
\end{figure}

It depends by the mixing angle $\alpha$, and the oscillation length $L=4\pi E/\Delta m^2$, where $E$ is the energy and $\Delta m^2\equiv \mid m_1^2-m_2^2\mid$ is the squared mass difference between the two mass eigenstates. By taking into account that
$\Delta m^2_{21}=(7.50\pm0.20)\times10^{-5}\text{eV}^2$, and $\Delta m_{32}^2=2.32^{+0.12}_{-0.08}\times10^{-3}\text{eV}^2$ \cite{alpha},
we can assume that the oscillation length $L$ is small compared to the linear dimension of the source, so that the source is flavor coherent and the oscillations will be averaged out both over dimension and energy. In any case, due to the large distance of the GC with respect to the dimensions of the detector, this fact does not affect the computation \cite{NeutrinoOsci}. For a point-like source localized in the GC, we can assume that the totally averaged oscillations among the three flavors is given by a symmetric matrix of the form:

\begin{equation}
\left(
\begin{array}{c}
\Phi_{\nu_e}\\
\Phi_{\nu_\mu}\\
\Phi_{\nu_\tau}
\end{array}
\right)
=
\left(
\begin{array}{c c c}
P_{ee} & P_{e\mu} & P_{e\tau} \\
P_{e\mu} & P_{\mu\mu} & P_{\mu\tau} \\
P_{e\tau} & P_{\mu\tau} & P_{\tau\tau}
\end{array}
\right)
\left(
\begin{array}{c}
\Phi^0_{\nu_e}\\
\Phi^0_{\nu_\mu}\\
\Phi^0_{\nu_\tau}
\end{array}
\right)\,.
\label{matrix}
\end{equation}

The elements $P_{\alpha\beta}$ depend on the three mixing angles $\alpha_{ij}$ and the CP phase $\delta$ (read, for example, \cite{Lai}). There are
important uncertainties associated to these values, but a good and simple approximation is given by assuming  $\text{sin}^2(2\alpha_{13})=0$ and $\text{sin}^2(2\alpha_{23})=1$ (the present experimental observations constraint these angles as $\text{sin}^2(2\alpha_{13})=0.095\pm0.010$ and $\text{sin}^2(2\alpha_{23})>0.95$ \cite{alpha}). In such a case, $P_{\alpha\beta}$ depends only on the $\alpha_{12}$ angle in the following way:
$P_{ee}\simeq 1- \text{sin}^2(2\alpha_{12})/2$, $P_{e\mu}\simeq P_{e\tau}\simeq 1- \text{sin}^2(2\alpha_{12})/4$,
$P_{\mu\mu}\simeq P_{\mu\tau}\simeq P_{\tau\tau}\simeq 1- \text{sin}^2(2\alpha_{12})/8$.

\begin{table*}[t]
\begin{center}
\begin{tabular}{|c||c|c|c|}\hline
\multirow{2}{*}{\backslashbox{\,\,\,\,\,$\theta^\circ$}{$E_{min}^{\nu}$(GeV)} }&\multicolumn{3}{c|}{$t_{exp}$}\\  \cline{2-4}
 & 2\,yr & 3\,yr &5\,yr \\
\hline
\hline
$5\sigma$ &\backslashbox{\,\,\,\,\,$0.18$}{$818$\,\,} & \backslashbox{\,\,\,\,\,$0.15$}{$630$\,\,}  &  \backslashbox{\,\,\,\,\,$0.23$}{$973$\,\,}\\
\hline
$3\sigma$   & \backslashbox{\,\,\,\,\,$0.24$}{$977$\,\,}  & \backslashbox{\,\,\,\,\,$0.32$}{$1102$\,\,}  & \backslashbox{\,\,\,\,\,$0.45$}{$1737$\,\,} \\
\hline
$2\sigma$  &\backslashbox{\,\,\,\,\,$0.42$}{$1321$\,\,}  &\backslashbox{\,\,\,\,\,$0.54$}{$1482$\,\,}  & \backslashbox{\,\,\,\,\,$0.72$}{$1811$\,\,} \\
\hline
\end{tabular}
\caption{Energy threshold cut (GeV) and resolution angle in order to achieve a confidence level of $5\sigma$, $3\sigma$ or $2\sigma$
from the muon neutrino flux for three different exposition times for DM annihilating into the $W^+W^-$ channel with an effective area of
$50\,\text{m}^2$.}
\label{Enmin1}
\end{center}
\end{table*}

\begin{table*}[t]
\begin{center}
\begin{tabular}{|c||c|c|c|}\hline
\multirow{2}{*}{\backslashbox{\,\,\,\,\,$\theta^\circ$}{$E_{min}^{\nu}$(GeV)} }&\multicolumn{3}{c|}{$t_{exp}$}\\  \cline{2-4}
 & 2\,yr & 3\,yr &5\,yr \\
\hline
\hline
$5\sigma$ & & \backslashbox{\,\,\,\,\,$0.002$}{$21$\,\,}  &  \backslashbox{\,\,\,\,\,$0.003$}{$156$\,\,}\\
\hline
$3\sigma$   & \backslashbox{\,\,\,\,\,$0.02$}{$110$\,\,}  & \backslashbox{\,\,\,\,\,$0.03$}{$176$\,\,}  & \backslashbox{\,\,\,\,\,$0.07$}{$334$\,\,} \\
\hline
$2\sigma$  &\backslashbox{\,\,\,\,\,$0.06$}{$296$\,\,}  &\backslashbox{\,\,\,\,\,$0.15$}{$638$\,\,}  & \backslashbox{\,\,\,\,\,$0.15$}{$624$\,\,} \\
\hline
\end{tabular}
\caption{Same data reported in Tab. \ref{Enmin1} but for an effective area of $5\,\text{m}^2$.}
\label{Enmin2}
\end{center}
\end{table*}

It means that the astrophysical flux of $\nu_\mu$ and $\nu_\tau$ are approximately the same independently of the flavor composition of neutrinos produced
at the source. In addition, as the value of $\alpha_{12}$ is important ($\text{sin}^2(2\alpha_{12})=0.857\pm0.024$ \cite{alpha}), the oscillation effects need
to be taking into account. In any case, as it can be seen in Figures \ref{MCfluxes} and \ref{EXPflux} for the $W^+W^-$ annihilation channel, we have checked
that the neutrino flavor ratio of the fluxes observed at the Earth are very homogeneous: $\Phi_{\nu_e}:\Phi_{\nu_\mu}:\Phi_{\nu_\tau}
\simeq 1:1:1$.
The reason is that the most part of the neutrinos come from the charged pion decay chain:
$\pi^+\rightarrow\mu^++\nu_\mu\rightarrow e^++\nu_\mu+\nu_e+\bar\nu_\mu$ (or $\pi^-\rightarrow\mu^-+\bar\nu_\mu\rightarrow e^-+\bar\nu_\mu+\bar\nu_e+\nu_\mu$),
that gives an original ratio: $\Phi^0_{\nu_e}:\Phi^0_{\nu_\mu}:\Phi^0_{\nu_\tau}\simeq 1:2:0$. This production is dominant except for the mentioned
$W^+W^-$ channel at very high energies, where the neutrinos are produced directly by the leptonic decay of the gauge bosons:
$W^+\rightarrow l^++\nu_l$ (or $W^-\rightarrow l^-+\bar\nu_l$), but it implies that even the original neutrino flux produced by the source is already homogeneous: $\Phi^0_{\nu_e}:\Phi^0_{\nu_\mu}:\Phi^0_{\nu_\tau}\simeq 1:1:1$.
In both cases, it is easy to understand from the oscillation Matrix (\ref{matrix}) that the three flavors arrive at the Earth with very similar fluxes.

The differential number of neutrinos for the different flavors $\nu_p$, with $p=e$, $\mu$ and $\tau$, as generated by the Monte Carlo event generator software, are shown in Fig. \ref{MCfluxes}. The photon differential number is also shown for reference. As we have commented, the three flavors are produced with the same ratio at high energies, whereas the number of $\nu_\tau$ is negligible at low ones. In Fig. \ref{EXPflux}, we show the expected neutrino fluxes
given by Eq. (\ref{nuflux}), as observed at the Earth, when oscillations and detection limits are taken into account. The parameters are given by the DM model independent fit of the HESS data in gamma-rays characterized by Eq. (\ref{astroneutrinoWW}) and $M\simeq 48.8$ TeV \cite{HESSfit}. At this stage, the energy resolution of the neutrinos detector has not been yet considered.

As we have commented, we are assuming that the neutrino detector will not be able to distinguish between neutrinos and antineutrinos  \cite{nue}.
So the neutrino flux $\Phi_{\nu_\alpha}$ is understood to be the sum of $\nu_\alpha$ and $\bar\nu_\alpha$. In addition, we shall assume that the detector will be able to distinguish muon neutrinos from electron and tau neutrinos. The later flavors give a typical showering signal, whereas the $\nu_{\mu}$ provide a distinctive track signal. 
More precisely, neutrino flavors can be deduced from two different event topologies: muon tracks, related to
the Cherenkov light of a propagating muon, and hadronic or electromagnetic showers.
Showers are produced by neutral current (NC) interactions of any neutrino flavor, and 
by both $\nu_e$ and $\nu_\tau$ charge current (CC) interactions. On the contrary, tracks are induced by muons 
from $\nu_\mu$ CC interactions and $\nu_\tau$ CC interactions in which the tau decay produces a muon.

\section{Analysis}

The most important source of background for highly energetic astrophysical neutrinos is given by atmospheric neutrinos and muons, depending on the direction of observation. The $\nu_\mu$ and $\nu_e$ atmospheric neutrinos have been reported by IceCube \cite{numu, nue}. The electronic neutrino background has few data with important uncertainties. In this case, the $\nu_e$ atmospheric flux can be well fitted by a simple power-law:
\begin{eqnarray}
E^2\times \frac{d \Phi^{Bg}_{\nu_e}}{dE}=A_e \left(\frac{E}{\text{GeV}}\right)^{-B_e^0},
\label{Bge}
\end{eqnarray}
with $A_e=0.012\pm 0.011$ GeV$\,$cm$^{-2}$s$^{-1}$sr$^{-1}$ and $B_e^0=1.17$. IceCube has measured the muon neutrino background with more detail, and a modified power-law fitting function is
needed to reproduce the observed data accurately:
\begin{eqnarray}
E^2\times \frac{d \Phi^{Bg}_{\nu_\mu}}{dE}=A_\mu \left(\frac{E}{\text{GeV}}\right)^{-(B_{\mu}^0+B_\mu\times ln(E/\text{GeV}))},
\label{Bgmu}
\end{eqnarray}
with $A_\mu=0.05^{+0.01}_{-0.02}$ GeV$\,$cm$^{-2}$s$^{-1}$sr$^{-1}$, $B_{\mu}^0=0.81^{-0.066}_{+0.008}$, and $B_\mu=0.037$.
The IceCube experimental data and both fitting functions within $1\sigma$ standard deviation are shown in Fig.\ref{WFigRIS}.
The lack of $\nu_e$ atmospheric flux data and its large uncertainty allow the power law fit, but a decreasing flux similar to the
$\nu_\mu$ case is expected at energies higher than $10^4$ GeV. As we shall discuss, the analysis associated with the $\nu_e$
signal is not particularly interesting in this case due to its lower angular accuracy. Therefore, the overestimation of its
atmospheric background at high energies does not have consequences in our results.

Our purpose is to estimate the possibilities of a general neutrino telescope to be sensitive to the neutrino signal associated
to the HESS observation by assuming a DM origin. In order to be conservative, we will consider a  $5\sigma$ signal
(or a less restrictive $3\sigma$ or $2\sigma$ confidence level) by comparing the number of events with respect to the atmospheric
background for a particular neutrino signature:
\begin{equation}
\chi_{\nu_i}=\frac{\Phi_{\nu_i}\sqrt{A_{\text{eff}}\,t_{\text{exp}}\,\Delta\Omega}}{\sqrt{\Phi_{\nu_i}+\Phi^{\text{Atm}}_{\nu_i}}} = 5\, (3,\, 2)\,,
\label{Chi1}
\end{equation}
where the effective area $A_{\text{eff}}$, the solid angle $\Delta\Omega$ and the exposition time $t_{\text{exp}}$ depend on the
particular detector and the observation. High energy neutrino telescopes have an effective area range between the cm$^2$ and the km$^2$,
depending not only on the experiment, but also on the neutrino energy, the position of the source with respect to the telescope and
the associated type of background.
We can combine the track search and the shower signals in a common analysis.
However, high energy muons point essentially in the same direction as the incident neutrino, and the angular
resolution of high energy muon tracks is quite good, smaller than $\theta= 1^\circ$ for detectors as IceCube.
This feature makes these signatures particularly interesting for the analysis of DM annihilation in the GC.
For the IceCube/DeepCore detector, the GC is above the horizon, so the neutrino flux from this region contributes to the
downward muon rate. However, for ANTARES \cite{ANTARES} or the proyected KM3NeT \cite{Km3} detector, the GC contributes to the upward muon rate.
This fact is a clear advantage since the effective area and volume are enhanced.

\begin{figure}[t]
\begin{center}
\epsfxsize=13cm
\resizebox{8cm}{8cm}
{\includegraphics{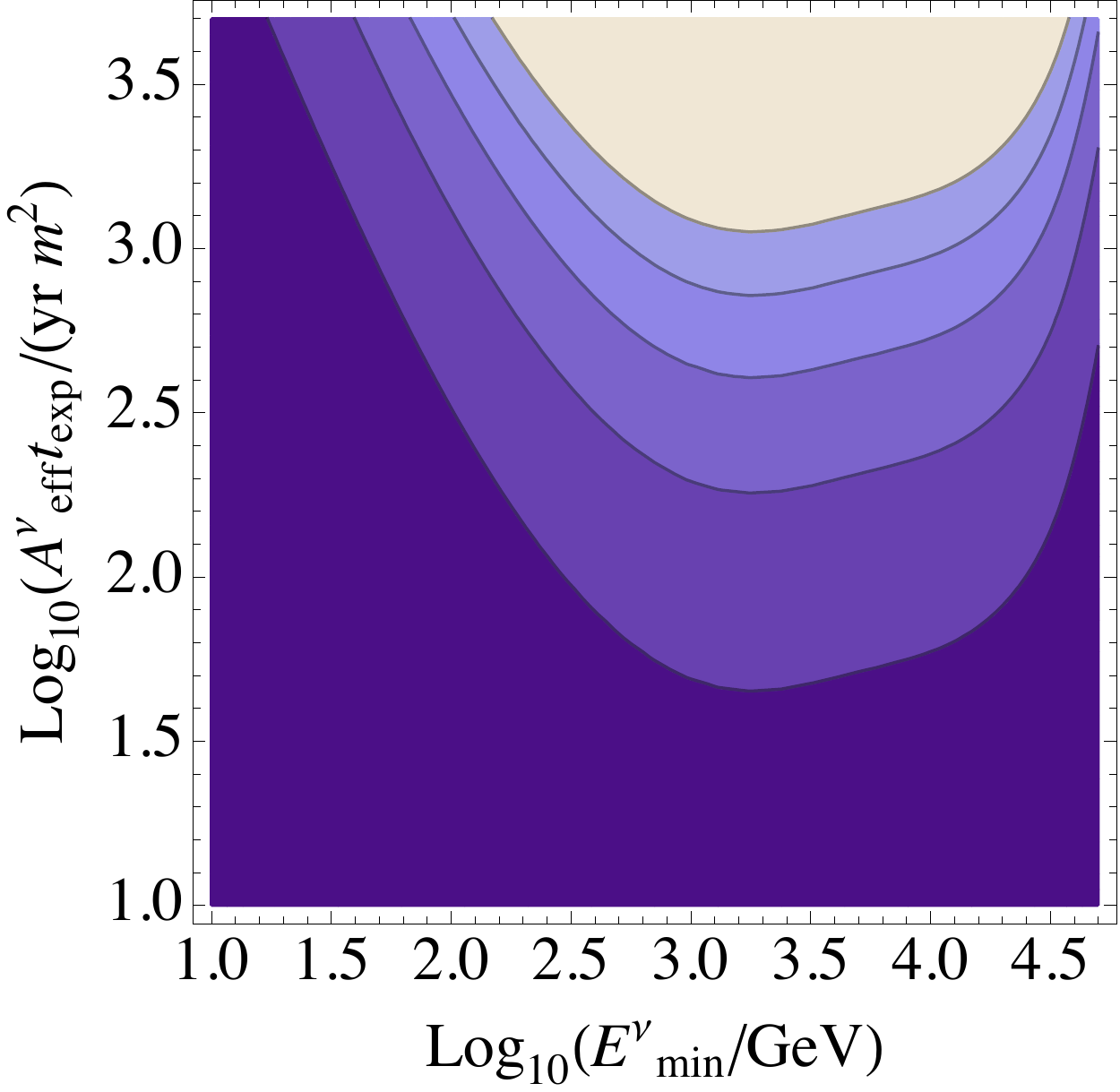}}
\end{center}
\begin{center}
\epsfxsize=13cm
\resizebox{8cm}{8cm}
{\includegraphics{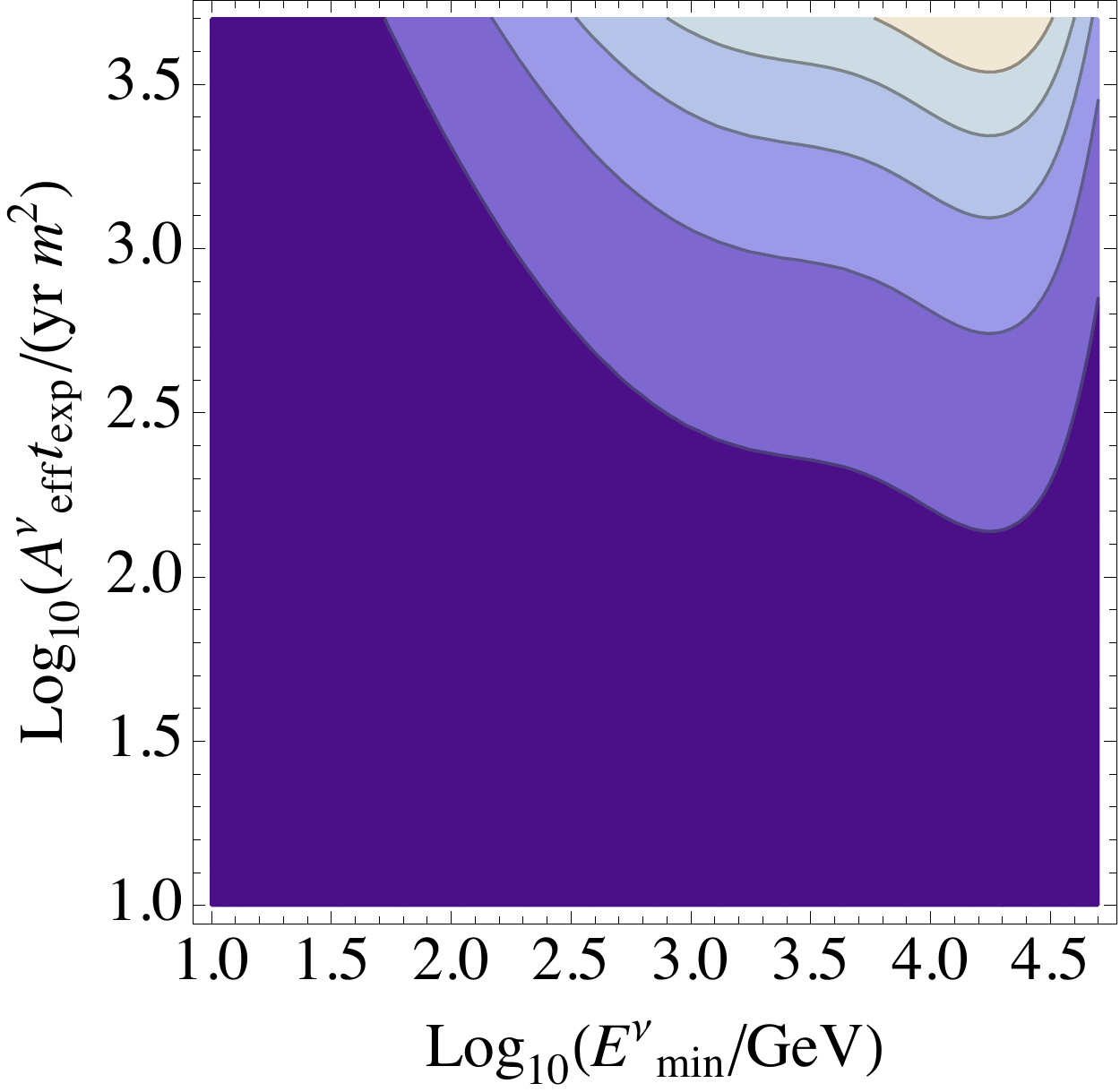}}
\caption {\footnotesize{
As in Figure \ref{thetaW}, the $1\sigma$ (dark), $2\sigma$, $3\sigma$, $4\sigma$, $5\sigma$ (white) confidence levels contours for
DM annihilating into the $W^+W^-$ channel are plotted. In this case, the angular field of view is fixed as
$\theta=0.6^\circ$ (top pannel) and $\theta=1.5^\circ$ (bottom pannel).
Therefore, the possibility to detect the neutrino flux signal above the atmospheric background depends on the energy cut $E_{min}^\nu$
and the factor $Af\equiv A_{\text{eff}}\times t_{\text{exp}}$.}}
\label{AtW}
\end{center}
\end{figure}

The electromagnetic or hadronic showers produced by neutrinos can be used as an additional signature to test the
DM interpretation of the muon track signal. However, it is difficult to think that they can be used to have the first
evidence of DM neutrinos coming from the GC since the current capabilities for shower angular resolution are much more limited.

As it can be observed in Fig. \ref{WFigRIS}, the sensitivity to DM in the GC depends crucially on the angular resolution.
The best strategy consists in reducing the angle in order to decrease the atmospheric background. In such a case, an
excess at energies of the order of $\sim 10$ TeV can be observable. In order to estimate the energy cut-off $E_{min}^\nu$
,
we can restrict the total background to few events: $\sum_{i=1}^2 \Phi^{\text{Atm}}_{\nu_i}\times A_{\text{eff}}\,t_{\text{exp}}\simeq 1$.
As we have commented, we will assume that neutrinos produced by a point-like source are independent on the resolution angle
of the neutrino telescope. In order to compute the number of neutrino events coming from DM, we integrate Eq. (\ref{nuflux}) over the
observation time and energy:

\begin{equation}
N^{t_{exp}}_{\nu_f}=\int^\infty_{E_{\text{min}}^\nu}dE_{\nu}\,\,\, \frac{d \Phi_{\nu_f}}{dE} \times A_{\text{eff}}\,t_{\text{exp}}\,.
\label{nevent}
\end{equation}

\begin{table*}
\begin{center}
\begin{tabular}{|c||c|c|c|}\hline
\multirow{2}{*}{\backslashbox{\,\,\,\,\,$\theta^\circ$}{$E_{min}^{\nu}$(GeV)} }&\multicolumn{3}{c|}{$t_{exp}$}\\  \cline{2-4}
 & 2\,yr & 3\,yr &5\,yr \\
\hline
\hline
$5\sigma$ & \backslashbox{\,\,\,\,\,$0.13$}{$274$\,\,}  & \backslashbox{\,\,\,\,\,$0.16$}{$336$\,\,}  &  \backslashbox{\,\,\,\,\,$0.22$}{$420$\,\,}\\
\hline
$3\sigma$   & \backslashbox{\,\,\,\,\,$0.24$}{$398$\,\,}  & \backslashbox{\,\,\,\,\,$0.30$}{$479$\,\,}  & \backslashbox{\,\,\,\,\,$0.40$}{$524$\,\,} \\
\hline
$2\sigma$  &\backslashbox{\,\,\,\,\,$0.38$}{$490$\,\,}  &\backslashbox{\,\,\,\,\,$0.46$}{$839$\,\,}  & \backslashbox{\,\,\,\,\,$0.60$}{$552$\,\,} \\
\hline
\end{tabular}
\caption{Same data reported in Tab. \ref{Enmin1} but in the case of DM annihilating into $u\bar u$ channel with an effective area of $50\,\text{m}^2$.}
\label{uEmin}
\end{center}
\end{table*}

\begin{figure}[b!]
\begin{center}
\epsfxsize=13cm
\resizebox{8.8cm}{6.6cm}
{\includegraphics{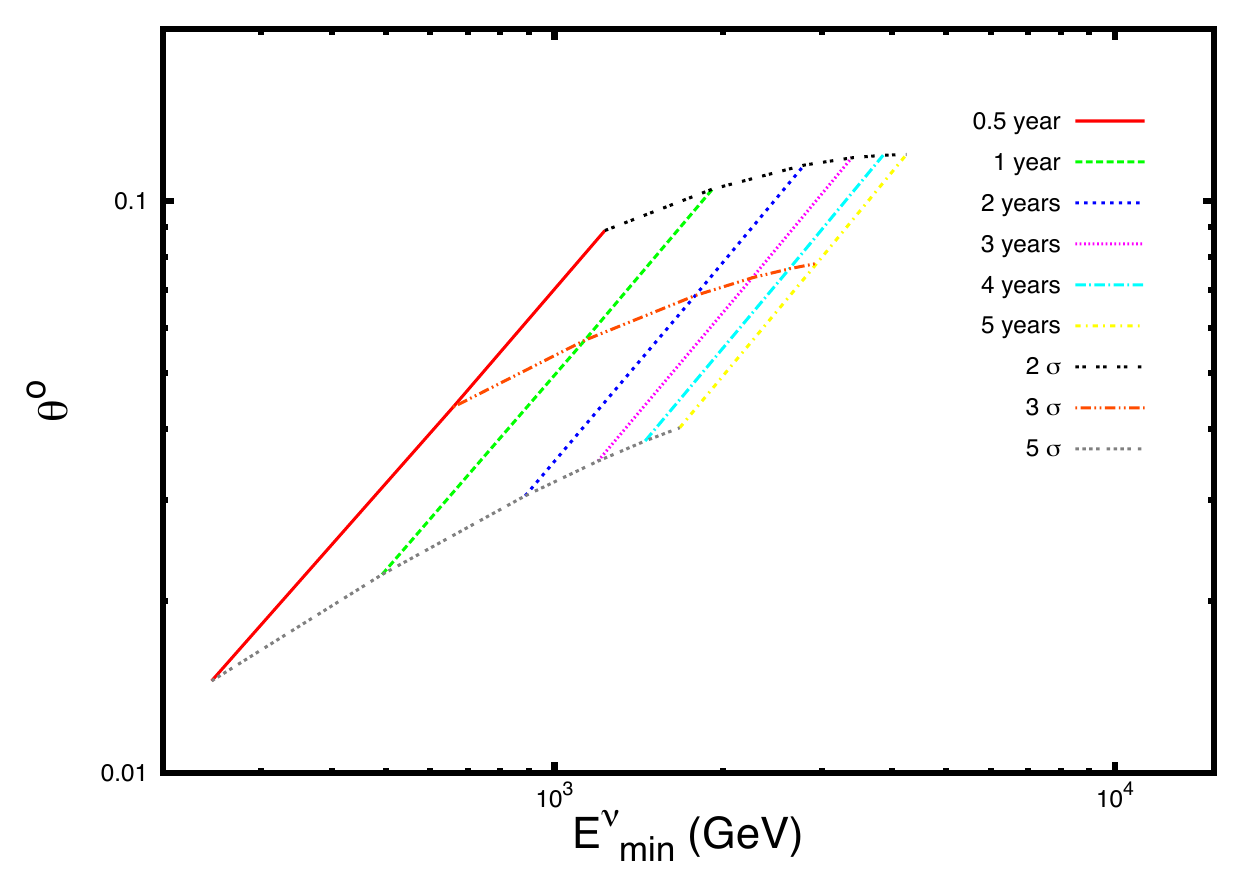}}
\caption {\footnotesize{
Same information as Fig. \ref{tfixRIS} but for the $u\bar u$ channel. In this case, the DM mass is fixed to $27.9$ TeV,
and the astrophysical factor is given by Eq. (\ref{astroneutrinouub}). The lowest value of $\theta\simeq 0.01^\circ$
corresponds to a $5\sigma$ confidence level
with energy threshold of $E^\nu_{\text{min}}\simeq 244$ GeV and six months of exposition time.
The largest value of $\theta\simeq 0.12^\circ$ is associated with $5$ years of
exposition time, a statistical significance of $2\sigma$, and an energy threshold of $E^\nu_{\text{min}}\simeq 4.25$ TeV.
}}
\label{utfixRIS}
\end{center}
\end{figure}

We shall not consider the probability to detect a neutrino due to closeness of its production to the detector. There is also an
attenuation effect associated with neutrinos interactions within the Earth's volume \cite{crocker, GCnu}.
It only affects to up coming neutrinos and it shall be also neglected in our estimations. By fixing the exposition
time ($t_{exp}=0.5,\, 1,\, 2,\,3,\,4,\,5$ years in
Figures \ref{tfixRIS} and
\ref{utfixRIS}), we can determine the minimum energy $E_{min}^\nu$ that gives a certain number of neutrino events for
each observation time (in the same Figures: $N_{\nu_\mu}\simeq 25,\, 9,\, 4$, which are approximately associated with $5,\,3$ or $2\sigma$ if the background events are negligible).
On the contrary to the neutrino flux from DM, the events corresponding to the atmospheric background depend on the resolution angle of the telescope.
For a given energy cut $E_{min}^\nu$, we can find the maximum value for the angular field of view $\theta$ necessary to detect a negligible background
(We have allowed $1$ event of background for the reported values in
Figures \ref{tfixRIS} and \ref{utfixRIS}). We have developed this analysis for two channels qualitatively different:
$W^+W^-$ boson and $u\bar u$ quark-antiquark annihilation.

\begin{figure}[b!]
\begin{center}
\epsfxsize=13cm
\resizebox{8.8cm}{6.6cm}
{\includegraphics{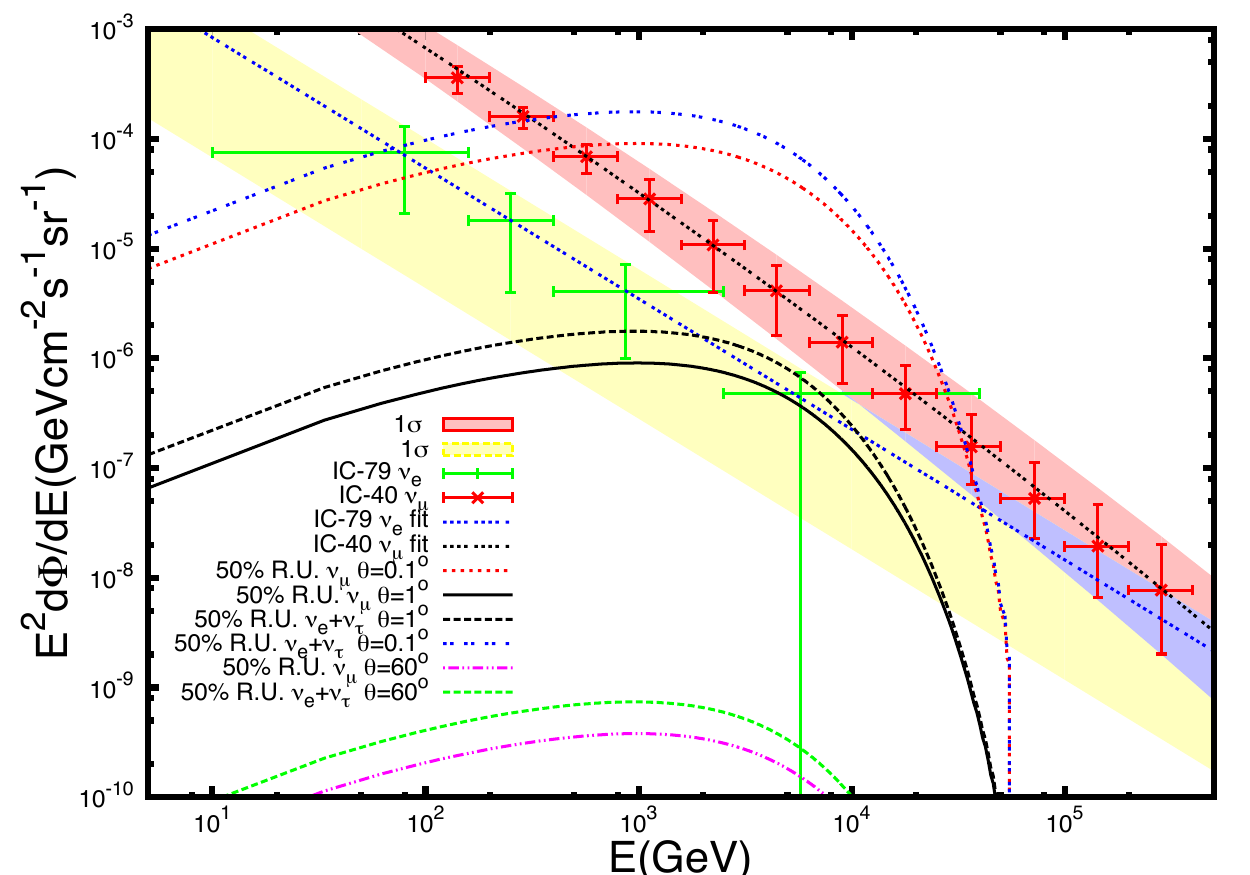}}
\caption {\footnotesize{Same information as Fig. \ref{WFigRIS} but for DM annihilating into the $u\bar u$ channel.}}
\label{uFigRIS}
\end{center}
\end{figure}

Following \cite{HESSfit}, DM annihilating into the $W^+W^-$ channel requests a DM mass of around $48.8$ TeV to fit the HESS gamma-ray spectra
of the J$1745$-$290$ source. As we can see in Fig. \ref{WFigRIS}, no neutrino signal produced by such kind of DM is expected with an angle of $\theta\approx60^\circ$. In the same figure, it is shown that the DM flux can be observable for $\theta\sim 1^\circ$ or smaller
(we are assuming a typical resolution energy of $50\%$).

On the other hand, Figures \ref{thetaW} and \ref{AtW} are plotted without any constraint in the number of background events.
The minimum energy thresholds for the $W^+W^-$ channel,
 are reported in Tables \ref{Enmin1}
and \ref{Enmin2}
 for different effective areas and exposition times.
We have studied the variation of the angular field of view and the energy cut. Larger sensitivities require very accurate angular resolutions.
An analysis of energies larger than $E_{min}^{\nu}\simeq 973$ GeV and an effective area of $A_{\text{eff}} \simeq 50$ m$^2$ with an
exposition time of $t_{\text{exp}}\simeq 5$ yr can provide $5 \sigma$ detection signal for angular resolutions of $\theta \simeq 0.23^\circ$.
Larger angular analyses of the order $\theta\simeq 0.7^\circ$ can provide first evidences of these signatures with less statistical
significance. In this case, the energy cut needs to be larger ($E_{min}^{\nu}\simeq 18$ TeV) in order to reduce the atmospheric background.
 In Fig. \ref{thetaW},
we show the resolution angle $\theta$ as function of the minimum energy cut $E_{min}^{\nu}$ for different statistical significances and
exposition times $t_{\text{exp}}$. Similar information about the factor $Af\equiv A_{\text{eff}} \times t_{\text{exp}}$ is given in Figure \ref{AtW}.

\begin{figure}[t!]
\begin{center}
\epsfxsize=13cm
\resizebox{8cm}{8cm}
{\includegraphics{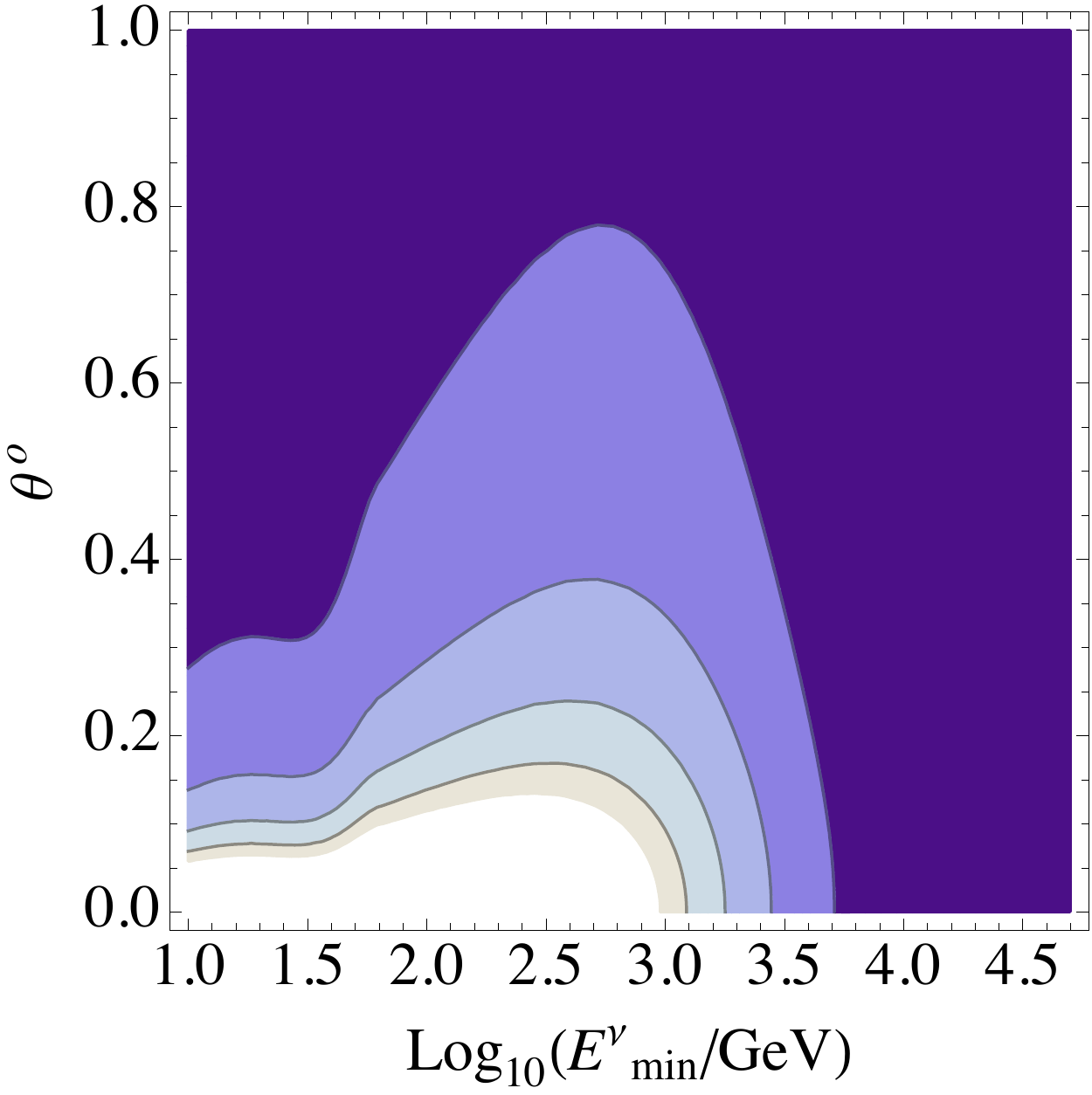}}
\end{center}
\begin{center}
\epsfxsize=13cm
\resizebox{8cm}{8cm}
{\includegraphics{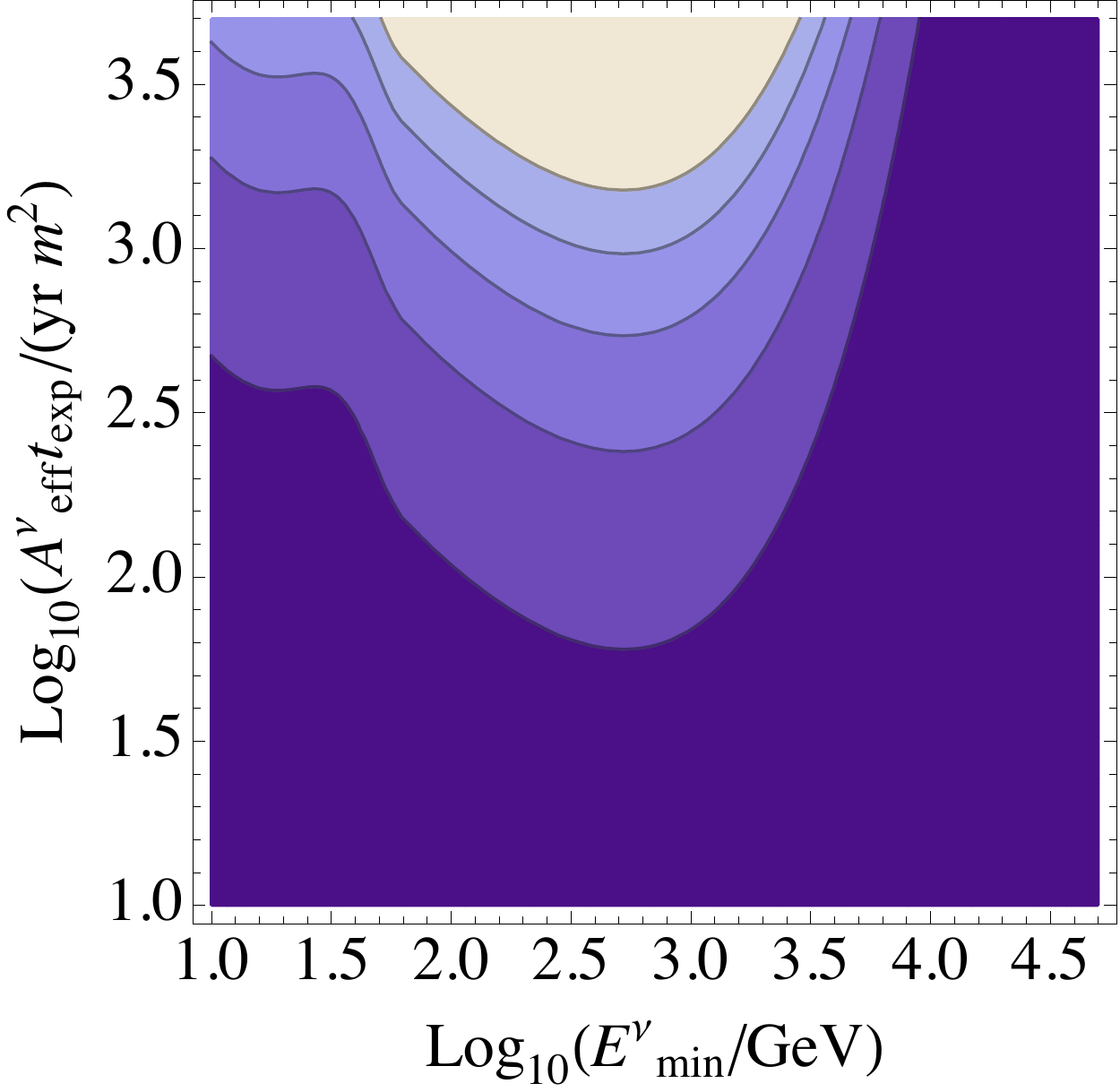}}
\caption {\footnotesize{ Confidence level contours associated to the observation of
DM annihilating into the $u\bar u$ quark-antiquark channel at $1\sigma$ (dark), $2\sigma$, $3\sigma$, $4\sigma$, $5\sigma$ (white) confidence level. Top panel :
The minimum energy cut is optimized around $1$ TeV depending on the resolution angle. The exposition time and effective area are fixed to the relation:
$Af\equiv A_{\text{eff}}\times t_{\text{exp}}\simeq 100\, \text{m}^2\,\text{yr}$.
 Bottom panel: the angular field of view is fixed as
$\theta=0.6^\circ$.
In such a case, the possibility to detect the neutrino flux signal above the atmospheric background demands
$Af\equiv A_{\text{eff}}\times t_{\text{exp}}\gtrsim 100\;\text{m}^2\text{yr}$.}}
\label{usigma}
\end{center}
\end{figure}

The J$1745$-$290$ gamma-rays spectrum observed by HESS can be also well fitted by DM annihilating in hadronic modes. As an example, we have analyzed
the $u\bar u$ quark-antiquark channel, which requires a mass close to $27.9$ TeV \cite{HESSfit}. Under this assumption, we have repeated the study
developed for the $W^+W^-$ channel. In Fig. \ref{uFigRIS}, we show the expected flux for different angular analyses. Estimations of the minimum
energy cut and resolution angles depending on the exposition time and the statistical significance with negligible background are reported in
 Fig. \ref{utfixRIS}. In Table \ref{uEmin} and Fig. \ref{usigma}, we present the results of the analysis for the same hadronic channel without constraining the number of background events, but fixing the effective area and exposition time combination ($Af=100\text{m}^2\text{yr}$ in the upper pannel) or the resolution angle
($\theta= 0.6^\circ$ in the lower panel).

\section{Conclusions}

The operation of the IceCube neutrino telescope at the South Pole, together with several counterparts at the Nothern hemisphere,
such as ANTARES and NT200 presently, or the future KM3NeT and GVD, are opening a new window in our knowledge of neutrino astronomy.

Indeed, the construction of KM3NeT will imply a new substantial improvement in sensitivity corresponding to a km$^3$ sized detector.
On the other hand, radio and airshower detectors, such as ANITA and the Pierre Auger observatory are sensitive to neutrinos with even
higher energies. The development of neutrino detectors have increased the interest for analysing the DM nature through the production
of astrophysical neutrinos as its primary source.

We have studied the prospective neutrino fluxes that should be originated by DM annihilating in the GC, in the case that the J1745-290
HESS high energy gamma-rays have this origin \cite{HESSfit}. The photon spectra is well fitted by different electroweak and hadronic
channels. We have done a explicit analysis for $48.8$ TeV DM annihilating in $W^+W^-$ and $27.9$ TeV DM annihilating into $u\bar u$ channel.
In these cases, the neutrino fluxes are completely determined by assuming that the DM region is localized as it is imposed by the gamma-rays
analysis. We have estimated the best combinations of energy cuts, observation times and angular resolutions of a general high energy neutrino
telescope.


For this purpose, we have used IceCube atmospheric neutrino observations as background. In particular, the data collected with exposition time of $t_{\text{exp}}^{\nu_\mu}=359$ days and $t_{\text{exp}}^{\nu_e}=281$ days for the muon and electron neutrinos, respectively \cite{numu, nue}.
We have found that for DM annihilating into the $W^+W^-$ boson channel, we need a resolution angle  $0.18^\circ\lsim\theta\lsim0.72^\circ$ and low
energy cut-off $818\,\text{GeV}\lsim E_{\text{min}}^\nu\lsim 1811$ GeV to get a signal between $5\sigma$ and $2\sigma$ with a minimum of 2 years
of exposition time and a maximum of five years for a $50\;\text{m}^2$ of detector effective area. The mass associated with the $u\bar u$ annihilation channel is significantly smaller. It implies that the neutrino flux produced in this case is less energetic, and more difficult to discriminate from the background. It demands a higher angular resolution ($0.13^\circ\lsim\theta\lsim0.60^\circ$) and the energy cuts need to be smaller ($274\,\text{GeV}\lsim E_{\text{min}}^\nu\lsim 552$ GeV) in order to
accumulate enough events. We have considered only track signal data by rejecting the muon background and taking into account the total number of events.
For a binned analysis with a non-zero background and with a combined analysis of track and shower signatures, it could be possible to find
better experimental configurations that should allow to detect neutrinos produced by heavy DM from the GC with worst resolution angle, smaller
effective area or less exposition time.

Recently, the IceCube collaboration have reported the observation of 28 high energy neutrinos over the range $30$ TeV $-\;1$ PeV at $4.1\sigma$ of confidence
level, and $t_{\text{exp}}=662\text{ days}$ $(\simeq1.8\text{ years})$. Of these events, 5 are likely originated from the GC \cite{GCnu}. These neutrinos seem to have an astrophysical origin, but the spectrum and direction are not compatible with the signal studied in this work (the angular resolution in the muon track events is of $\theta\approx8^\circ$). The DM signal analyzed in this work may only account for a small part of the events, that will be more likely associated with an electroweak channel, as the $W^+W^-$ annihilating DM model.

\vspace{0.5cm}

{\bf Acknowledgements}
We would like to thank Juande Zornoza and Carlos de los Heros for useful comments.
This work has been supported by UCM predoctoral grant, MICINN (Spain) project numbers FIS 2008-01323, FIS2011-23000, FPA2011-27853-01
and Consolider-Ingenio MULTIDARK CSD2009-00064.


\begin{thebibliography}{99}

\bibitem{CANG}
  K. Tsuchiya {\em et.~al.}, ApJ {\bf 606}, L115 (2004).

\bibitem{VER}
  K. Kosak {\em et.~al.}, ApJ {\bf 608}, L97 (2004).

\bibitem{MAG}
  J. Albert {\em et.~al.}, ApJ {\bf 638}, L101 (2006).

\bibitem{Vitale}
  V.~Vitale, A.~Morselli and f.~t.~F.~/L.~Collaboration, arXiv:0912.3828 [astro-ph.HE].

\bibitem{ferm}
  M. Cherenyakova {\em et.~al.}, ApJ {\bf 726}, 60 (2011);
  T.~Linden, E.~Lovegrove and S.~Profumo,  arXiv:1203.3539 [astro-ph.HE]. 

\bibitem{Aha}
  F. Aharonian {\em et.~al.}, A\&A {\bf 425},  L13 (2004).

\bibitem{HESS}
  F. Aharonian {\em et.~al.}, A\&A {\bf 503}, 817 (2009).

\bibitem{X}
  Q. Wang, F. Lu and E. Gotthelf, MNRAS {\bf 367}, 937 (2006);
  B. Aschenbach, N. Grosso, D. Porquet {\em et.~al.}, A\&A  {\bf 417}, 71 (2004).

\bibitem{Bergstrom1}
  L.~Bergstrom, T.~Bringmann, M.~Eriksson and M.~Gustafsson,  Phys.\ Rev.\ Lett.\  {\bf 94}, 131301 (2005);
  Phys.\ Rev.\ Lett.\  {\bf 95}, 241301 (2005).

\bibitem{DMint}
  F. Aharonian {\em et.~al.}, Phys. Rev. Lett. {\bf 97}, 221102 (2006).

\bibitem{HESSfit}
  J. A. R. Cembranos, V. Gammaldi, A.L. Maroto,  Phys. Rev. D 86, 103506 (2013); 
  JCAP {\bf 1304}, 051 (2013). 

\bibitem{Cohen}
  J. Cohen-Tanugi {\em et.~al.}, Proc. 31st ICRC (Lodz) 645 (http://icrc2009.uni.lodz.pl/proc/pdf/icrc0645.pdf)

\bibitem{SgrA}
  R.~M.~Crocker {\em et.~al.}, ApJ {\bf 622}, 892 (2005).

\bibitem{Atoyan}
  A. Atoyan and C. D. Dermer, ApJ {\bf 617}, L123 (2004).

\bibitem{AN}
  F. Aharonian and A. Neronov, ApJ {\bf 619}, 306 (2005).

\bibitem{Navarro:1996gj}
  J.~F. Navarro, C.~S. Frenk, and S.~D. White, ApJ {\bf 490}, 493 (1997).

\bibitem{Blumenthal}
  G.R. Blumenthal, S.M. Faber, R. Flores, J. R. Primack, ApJ {\bf 301}, 27 (1986);
  O.~Y.~Gnedin, A.~V.~Kravtsov, A.~A.~Klypin and D.~Nagai, ApJ  {\bf 616}, 16 (2004).

\bibitem{Prada:2004pi}
  F.~Prada, A.~Klypin, J.~Flix Molina, M.~Mart\'inez, E.~Simonneau, Phys.\ Rev.\ Lett.\  {\bf 93}, 241301 (2004).

\bibitem{Romano}
  E.~{Romano-D{\'{\i}}az}, I.~{Shlosman}, Y.~{Hoffman}, and C.~{Heller},  ApJ {\bf 685}, L105 (2008);
  ApJ {\bf 702}, 1250  (2009);
  A.~V. Maccio' {\em et.~al.}, arXiv:1111.5620 [astro-ph.CO].   

\bibitem{lab}
  J.~Alcaraz {\it et al.}, Phys. Rev. D {\bf 67}, 075010 (2003); 
  P. Achard {\it et al.}, Phys. Lett. {\bf B597}, 145 (2004); 
  J.~A.~R.~Cembranos, A.~Dobado and A.~L.~Maroto, Phys. Rev. {\bf D65} 026005 (2002); 
  Phys. Rev. {\bf D70}, 096001 (2004); 
  Phys.\ Rev.\ D {\bf 73}, 035008 (2006); 
  Phys.\ Rev.\ D {\bf 73}, 057303 (2006); 
  J.\ Phys.\ A  {\bf 40}, 6631 (2007); 
 J.~A.~R.~Cembranos, J.~L.~Diaz-Cruz and L.~Prado, Phys.\ Rev.\ D {\bf 84}, 083522 (2011). 
 J.~A.~R.~Cembranos, R.~L.~Delgado and A.~Dobado, Phys.\ Rev.\ D {\bf 88}, 075021 (2013). 


\bibitem{cosmics}
  J.~A.~R.~Cembranos, J.~L.~Feng and L.~E.~Strigari, Phys.\ Rev.\ Lett.\  {\bf 99}, 191301 (2007); 
  J.~A.~R.~Cembranos and L.~E.~Strigari, Phys.\ Rev.\  D {\bf 77}, 123519 (2008); 
  J. A. R. Cembranos, A. de la Cruz-Dombriz, V. Gammaldi, A.L. Maroto,  Phys. Rev. D 85, 043505 (2012). 

\bibitem{pythia}
  T. Sjostrand, S. Mrenna and P. Skands, JHEP05 (2006) 026 (LU TP 06-13, FERMILAB-PUB-06-052-CD-T) [hep-ph/0603175].

\bibitem{Ce10}
  J.~A.~R.~Cembranos, A. de la Cruz-Dombriz, A.~Dobado, R.~Lineros and A.~L.~Maroto,
  Phys.\ Rev.\  D {\bf 83}, 083507 (2011); 
  AIP Conf.\ Proc.\  {\bf 1343}, 595-597 (2011); J.\ Phys.\ Conf.\ Ser.\  {\bf 314}, 012063 (2011);
  A.~de la Cruz-Dombriz and V.~Gammaldi, arXiv:1109.5027 [hep-ph].

\bibitem{Cirelli}
M.Cirelli, G.Corcella, A.Hektor, G.HŸtsi, M.Kadastik, P.Panci, M.Raidal, F.Sala, A.Strumia, JCAP 1103 (2011) 051. 

\bibitem{NeutrinoOsci}
R. M. Crocker, F. Melia and R. R. Volkas, arXiv:9911292v2;
N.F. Bell,  arXiv:0811.0847v1.

\bibitem{Lai}
  K. C. Lai, G. L. Lin and T. C. Liu,
   Phys.\ Rev.\  D {\bf 80}, 103005 (2009). 

\bibitem{alpha}
J.Beringer et al. (Particle Data Group), PR D86, 010001 (2012).

\bibitem{nue}
M.G. Aartsen et al., IceCube Collaboration, arXiv:1212.4760v2 (2012).

\bibitem{numu}
R. Abbasi et al., IceCube Collaboration, arXiv:1010.3980v2 (2010);  M.~G.~Aartsen {\it et al.}  [IceCube Collaboration],
  Astrophys.\ J.\  {\bf 779} (2013) 132. 

\bibitem{ANTARES}
S. Adrian-Martinez et al. ANTARES Collaboration, Astrophy. J. 760:53(2012), arXiv:1207.3105 (2012); S. Schulte for the ANTARES Collaboration, icrc2013-0425.

\bibitem{Km3}
T. Seitz, R. Shanidze KM3NET Consortium, Nuclear Instrument and Methods in Physics Research A 626-627 (2011) S205-S207.

\bibitem{crocker}
R. M. Crocker, F. Melia, R. R. Volkas, arXiv:astro-ph: 0411471v4 (2003).

\bibitem{GCnu}
N. Whitehorn, C. Kopper, N.K. Neilson for the IceCube Collaboration at IceCube Particle Astrophysics Symposium 2013,
Madison, Wisconsin, USA; F. Halzen, S. Klein and C. Kopper, Proceedings of the 33th International Cosmic Ray Conference
(http://www.cbpf.br/icrc2013);
D. B. Fox, K. Kashiyama, P. Meszaros, arXiv:1305.6606v3 (2013);
S. Razzaque, arXiv:1309.2756v1, http://astro.fnal.gov/events/Seminars/Slides/
NWhitehorn061013.pdf.

\end{thebibliography}
\end{document}